\documentclass{iaus}
\usepackage{graphicx}

\title[Calibration synchrotron emission] 
{Measuring and calibrating\\
 Galactic synchrotron emission}

\author[W. Reich \& P. Reich]   
{Wolfgang Reich %
\and Patricia Reich}

\affiliation{Max-Planck-Institut f{\"u}r Radioastronomie, Auf dem H{\"u}gel 69,
52121 Bonn, Germany \break email: wreich, preich@mpifr-bonn.mpg.de\\[\affilskip]
}

\pubyear{2009}
\volume{259}  
\pagerange{100--100}
\date{"YOUR MAILING DATE"  and in revised form ??}
 \jname{Cosmic Magnetic Fields: From Planets,
> to Stars and Galaxies} \editors{K.G. Strassmeier, A.G. Kosovichev \&
> J.E. Beckman, eds.}

\begin{document}

\maketitle

\begin{abstract}
Our position inside the Galaxy requires all-sky surveys to reveal its large-scale properties.
The zero-level calibration of all-sky surveys differs from standard 'relative' measurements, where
a source is measured in respect to its surroundings. All-sky surveys aim to include emission
structures of all angular scales exceeding their angular resolution including isotropic emission
components.
Synchrotron radiation is the dominating emission process in the Galaxy up to frequencies of a few
GHz, where numerous ground based surveys of the total intensity up to 1.4~GHz exist.
Its polarization properties were just recently mapped for the entire sky at 1.4~GHz.
All-sky total intensity and linear polarization maps from WMAP for frequencies of 23~GHz
and higher became available and complement existing sky maps. Galactic plane surveys
have higher angular resolution using large single-dish or synthesis telescopes. Polarized diffuse emission
shows structures with no relation to total intensity emission resulting from Faraday rotation effects
in the interstellar medium. The interpretation of these polarization
structures critically depends on a correct setting of the absolute zero-level in Stokes U and Q.
\end{abstract}

\firstsection 

\section{Introduction}

All-sky radio continuum surveys provide basic information on our local environment and
the large-scale properties of the Galaxy. They are required to model the Galactic emission
components in 3-D (Sun {\it et al.} (2008)) and guide more sensitive higher-angular resolution
observations of the Galactic plane or other regions or objects of interest.
All-sky surveys are quite time consuming projects, which require special observing methods to
accurately measure large-scale sky emission. They need similar telescopes in the
northern and southern hemisphere and have to adapt calibration data from additional instruments
to provide the absolute level of sky emission. Thus all-sky surveys are rare, in particular in
linear polarization, where the signals are much weaker than in total intensities.

In the following we describe the basic methods to calibrate and adjust total intensity surveys as well
as the more complex requirements being applied for polarization surveys. We do not discuss the
instrumental corrections to be taken into account during the data reduction process, which can be found
in the original publications.

\section{Radio continuum measurements}\label{sec:radiocont}

When pointing a radio telescope to a certain sky direction we record - beside the signal
of interest - a variety of other much stronger signals. The observed signal can be expressed
as a temperature $\rm T_{obs}$:

\begin{equation}
  \rm T_{obs} = T_{sys} + T_{atm} + T_{ground} + T_{CMB} + T_{conf} + T_{gal} + T_{sou}
\end{equation}

where $\rm T_{sys}$ is the contribution from all components of the receiving system, typically
20~K to 30~K for a cooled receiver for cm-wavelength observations.
$\rm T_{atm}$ is the contribution  from the atmosphere, typically a few K, depending on elevation
and the actual weather conditions.
$\rm T_{ground}$ is the emission picked up by the sidelobes of a telescope from the ground, thus
depending on azimuth and elevation. $\rm T_{CMB}$ is the wavelength independent isotropic radiation
from the cosmic microwave background (CMB) of 2.73~K (Mather {\it et al.} (1994)). $\rm T_{conf}$ also is an isotropic
component resulting from unresolved weak extragalactic sources. $\rm T_{conf}$ depends on wavelength
and beam size. $\rm T_{gal}$ is the diffuse Galactic background emission and, finally, $\rm T_{sou}$ the
signal of interest from a specific Galactic or extragalactic source. $\rm T_{sou}$ and - in the case
of Galactic surveys - $\rm T_{gal}$ are to be separated from the other components in an
appropriate way. The problem is the weakness of the sky emission compared to
 the other contributions to $\rm T_{obs}$.

\subsection{Absolute versus relative measurements}

'Relative measurements' are the standard mode for continuum and polarization observations, when the
signal from a source or a certain region of the sky is of interest. In that case the region surrounding
the source or the source complex is refered to as the
zero-level and the difference signal is taken as $\rm T_{sou}$. This is justified in case the measurements
of the source signal and its adjacent zero-level are not suffering from time or telescope position dependent
changes of the unwanted contributions in an unpredictable way. Galactic diffuse emission shows intrinsic
variations, which might become a problem when they mix up with emission from extended faint sources,
e.g. a large diameter supernova remnant (SNR), and can not be separated. Thus large errors for integrated
flux densities reported for extended SNRs are very common.

An 'absolute measurement' means that the entire signal from the sky is recovered and the three
terms $\rm T_{sys} + T_{atm} + T_{ground}$ are subtracted. This is required for Galactic total intensity
and polarization all-sky surveys, but not easy to obtain  with high precision at decimeter and shorter
wavelengths. The two isotropic sky components $\rm T_{CMB}$ and $\rm T_{conf}$ don't matter in this context.
Special observing and reduction procedures are applied for survey observations. For instance, very long
scans across the sky are observed along the same azimuth or elevation range to ensure that  $\rm T_{atm}$
and/or $\rm T_{ground}$ are as constant as possible, while the sky emission varies as a function of time.
From an appropriate analysis of a large number of scans individual corrections for each scan against
the mean of all scans can be calculated. The final all-sky survey map is constructed from these
corrected scans. Nevertheless, the maps often show 'striping effects' along the scan direction at the noise
level or exceeding it, which indicate residuals from the correction procedures. In general
distortions of the measurements on short time scales, e.g. by weather changes or variing low-level RFI,
are more difficult to correct
than smooth changes of the baseline level by temperature effects over a night.

\begin{figure}
\centering
 \includegraphics[bb = 55 64 521 344,width=12cm,angle=0,clip=]{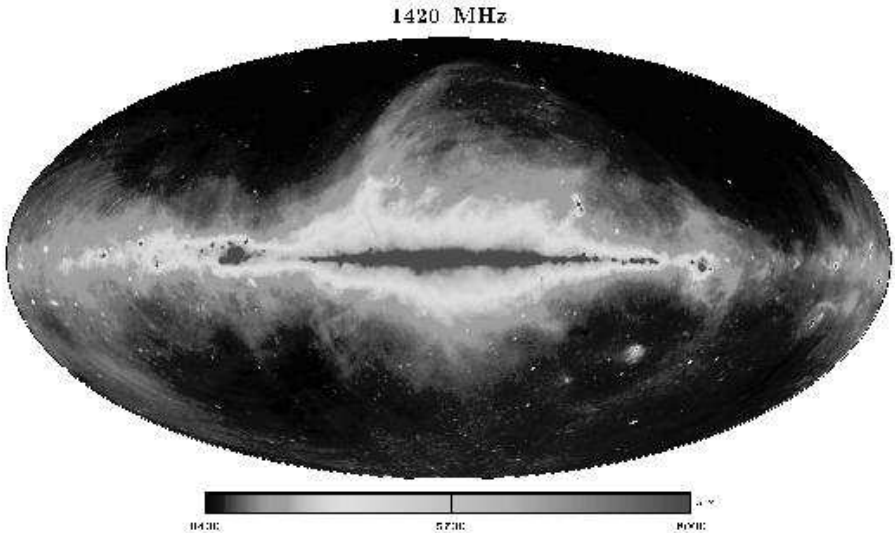}
  \caption{All-sky survey at 1.42~GHz combined from northern sky data observed with the Stockert 25-m
telescope near Bonn/Germany (Reich 1982; Reich \& Reich 1986)
and southern sky observations observed with a 30-m dish at Villa Elisa/Argentina
(Reich {\it et al.} (2001)).}\label{fig-all-I}
\end{figure}

\subsection{Sky horn measurements}

Continuum all-sky surveys receive their zero-level calibration by using low-resolution sky horn data.
Their antenna diagrams are well known to enable a proper far-sidelobe correction. Reference loads
at precisely known temperatures are required to find $\rm T_{sys}$. The all-sky surveys need to be convolved to the
beam size of the sky horns, which are typically several degrees, to find the temperature offset.
This way the 1.42~GHz survey (Fig.~1) was calibrated to an accuracy of 0.5~K~$\rm T_{B}$. The 408~MHz
survey by Haslam {\it et al.} (1982) has a zero-level accuracy of about 3~K~$\rm T_{B}$. This accuracy
is lower than the sensitivity (3~x~r.m.s-noise) of both surveys of 0.05~K~$\rm T_{B}$ and
2K~$\rm T_{B}$, respectively.

The probably most famous sky horn observations are those by Penzias \& Wilson (1965)
leading to the discovery of the 3~K cosmic microwave background radiation. This led to the award of the
Nobel Prize in 1978. Their absolute measurements revealed an isotropic sky component of 3.5~K,
interpreted by Dicke {\it et al.} (1965) to originate from the CMB. This 4.08~GHz measurement of 3.5~K happened
at a time, when 'relative' measurements in the mK~range were regularly made. This clearly reflects the
technical challenges for these kind of measurements.

\subsection{Adjusting surveys by TT-plots}

Ground based surveys carried out at different frequencies with their zero-level adapted from
sky horn measurements may be further adjusted relative to each other by using the so-called TT-plot
method according to Turtle {\it et al.} (1962). This method was discussed in some detail by Reich
{\it et al.} (2004).
The surveys to be adjusted were convolved to a common angular resolution of 15$^{\circ}$
and the TT-plots were performed for a strip in right ascension for declinations between 30$^{\circ}$
and 45$^{\circ}$, where no contamination from local structures like the giant loops
is evident. As an example we show
the TT-plot between the 408~MHz survey by Haslam {\it et al.} (1982) and the 1420~MHz survey by Reich (1982)
in Fig.~2 (left panel) after a zero-level correction of -2.7~K for the 408~MHz survey.
This correction is within its quoted zero-level error of 3~K. The mean of the fitted lines passes 0~K
at both wavelengths. This result assumes the 1420~MHz survey to be correct, although the sky horn
data are uncertain by 0.5~K. Using several pairs of surveys improves and constrains
the corrections (see Reich {\it et al.} (2004) for details).

The WMAP total intensity all-sky surveys provide valuable high-frequency maps of Galactic emission and are of
similar low-angular resolution compared to the ground based surveys up to 1.4~GHz. Unfortunately
the WMAP total intensity maps have not yet set on an absolute zero-level. The three year's release
(Hinshaw {\it et al.} (2007)) at 22.8~GHz shows numerous small patches at high latitudes with temperatures
below -100~$\mu$K, which is incorrect. We used the TT-plot technique to adjust the WMAP
22.8~GHz map in respect to 1420~MHz (see Fig. 2, right panel) and find an offset of 250$\pm70~\mu$K.
For the recent WMAP release of five years observations (Hinshaw {\it et al.}, preprint), an offset of
260$\pm70~\mu$K using the same method. This is quite a significant correction
and has a clear effect on the spectral index between 1420~MHz and 22.8~GHz. The most frequent high-latitude
temperature spectral indices increase from about -3.1 (without correction)
to about -2.8, which is just slightly
steeper (by about 0.1) compared to the spectra between 408~MHz and 1420~MHz.

\begin{figure}
\centering
 \includegraphics[width=6.5cm,angle=0]{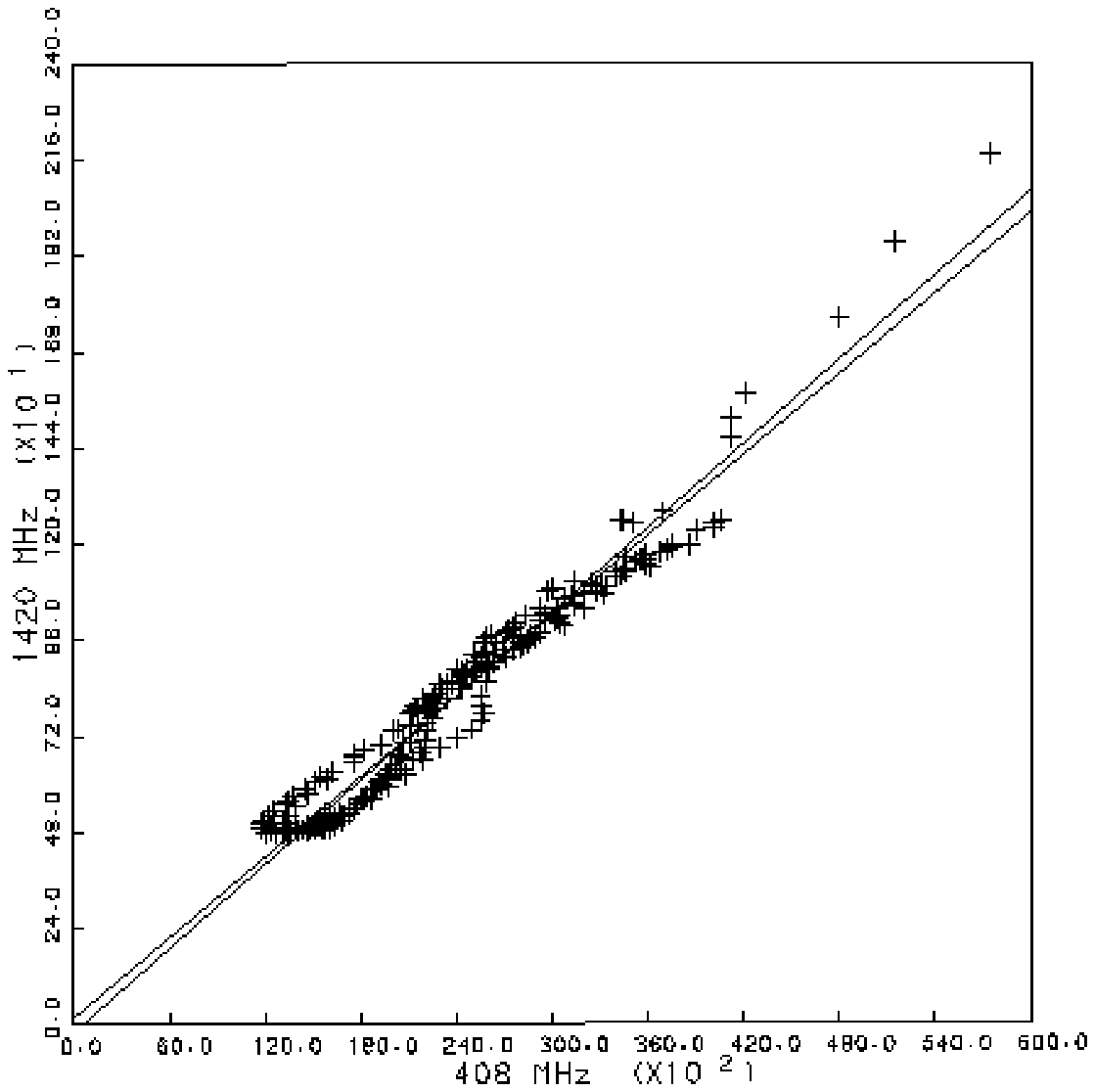}
 \includegraphics[width=6.5cm,angle=0]{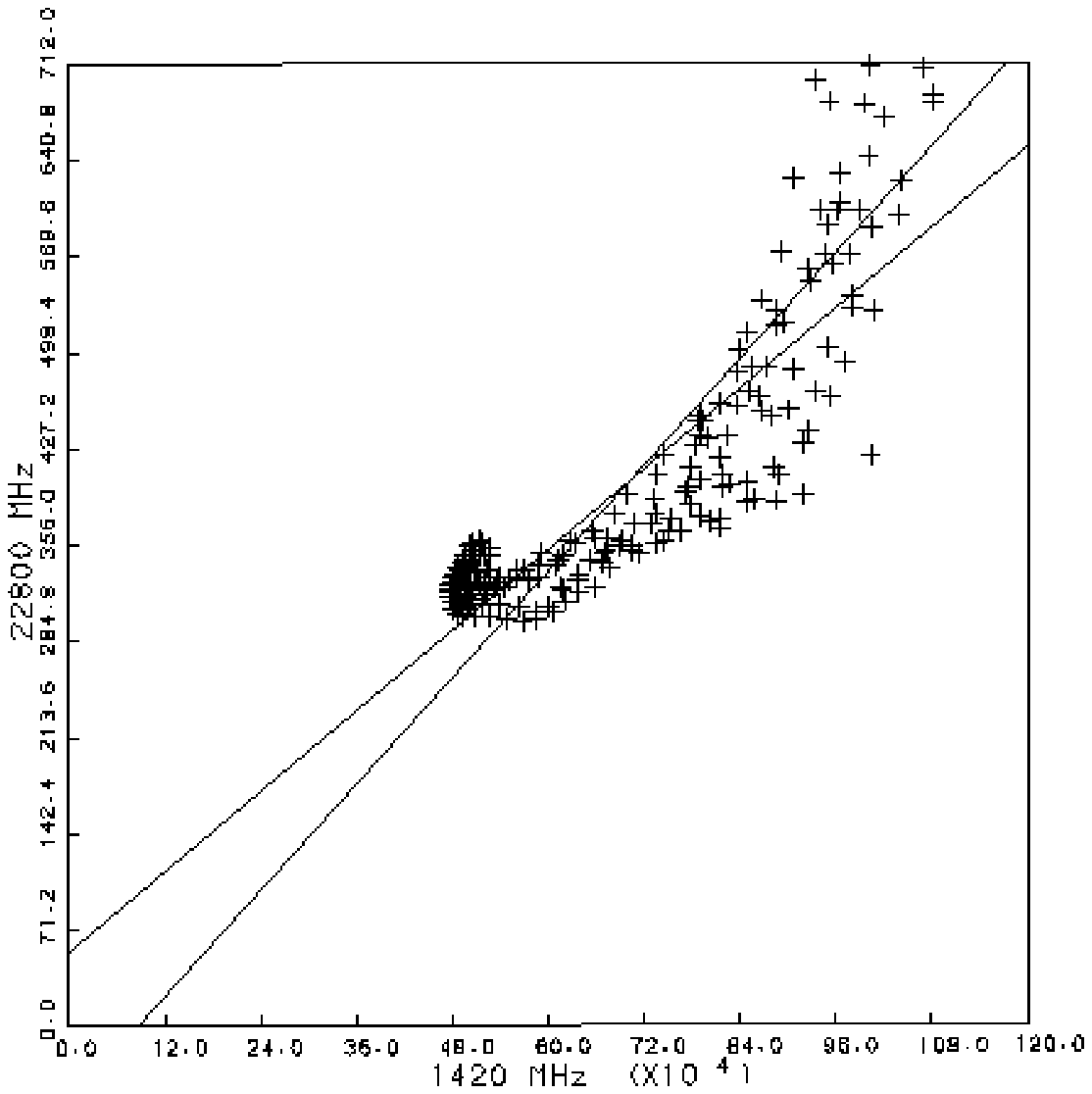}
  \caption{Left panel: TT-plot between the 408~MHz and 1420~MHz total intensity surveys leading to a
zero-level correction of -2.7~K for the 408~MHz survey. Right panel: TT-plot between 1420~MHz
and 22.8~GHz leading to an offset correction of the WMAP 3-yr survey by +250~$\mu$K.}\label{fig-TT}
\end{figure}

Absolute sky measurements with high precision are needed to calibrate the WMAP total intensity maps.
This is expected from ARCADE (Kogut {\it et al.} (2006)), a balloon-borne instrument designed to measure
the absolute sky temperature between 3.3~GHz and 90~GHz in a number of channels. This exeriment will
provide high-precision low-resolution data of Galactic emission.

\subsection{Galactic plane surveys}

Current Galactic plane surveys were carried out with large single-dish telescopes at arcmin angular resolution
to resolve diffuse Galactic emission structures from individual sources like SNRs or HII-regions.
Synthesis telescope surveys achieve even higher angular resolutions. These surveys use all-sky surveys
to add the missing large-scale information and adjust their zero-level. This has been done using the
total intensity northern sky 1.4~GHz survey
(Fig.~1) to adjust 1.4~GHz Effelsberg total intensity surveys (Kallas \& Reich 1980; Reich {\it et al.} (1990,
1997); Uyan{\i}ker {\it et al.} (1999)). These combined data at about 9' angular resolution
were used to calibrate the Canadian Galactic Plane
Survey (CGPS) carried out with the synthesis telescope at DRAO/Penticton as described by Taylor {\it et al.} (2003),
thus providing a survey including all structural information larger than 1'.
Similarly the CGPS survey at 408~MHz uses the all-sky survey carried out by Haslam {\it et al.} (1982) with the
Effelsberg, Parkes and Jodrell Bank telescopes at that frequency.

\section{Calibration of polarization data}\label{sec:pol}

The scheme for 'relative' and 'absolute' zero-level calibration of polarization data, which means the
calibration of the observed Stokes U and Q maps, in principle follows that of total intensity data:
The absolute zero-level is required for all-sky maps and a relative zero-level setting is sufficient for
polarized sources. The only ground-based all-sky polarization survey was made at 1.4~GHz by Wolleben
{\it et al.} (2006) for the northern hemisphere using the DRAO 26-m telescope and by Testori {\it et al.}
(2008) using one of the Villa Elisa 30-m telescopes for the southern sky, where
the different calibration steps are discussed. The 'absolute zero-level' of this polarization survey
is adapted to 1.4~GHz measurements with the Dwingeloo 25-m telescope published by Brouw \& Spoelstra (1976),
where rotating dipoles were used allowing to separate sky emission from ground radiation contamination.
The polarization intensity scale of these surveys, however, was adjusted by using smoothed polarization
data from the Effelsberg or the Parkes telescopes, respectively.
We show the combined all-sky 1.4~GHz polarization survey (Reich {\it et al.}, in prep.) in Fig.~3.

\begin{figure}
\centering
 \includegraphics[bb = 26 28 528 351,width=12cm,angle=0,clip=]{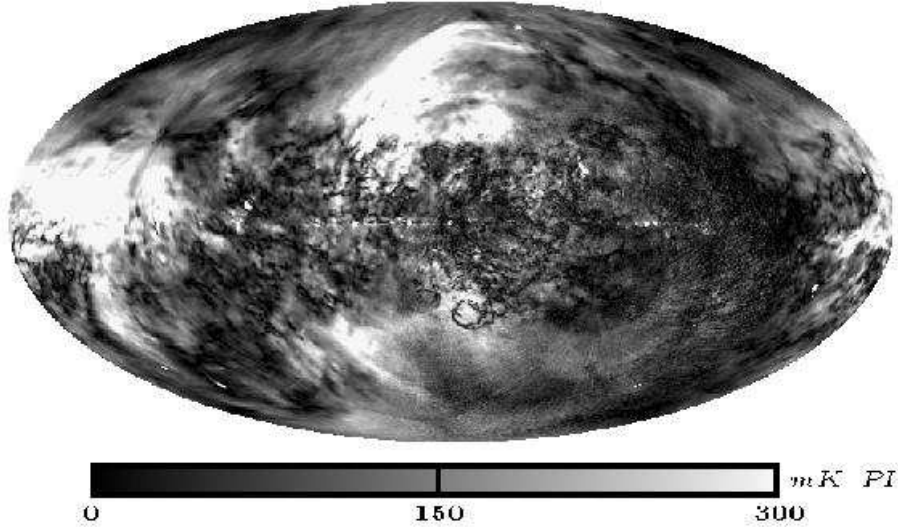}
  \caption{1.4~GHz polarization all-sky survey combined from northern sky observations by Wolleben et al.
(2006) and southern sky observations by Testori et al. (2008) at an absolute zero-level.
The angular resolution of the survey is about 36'
and its sensitivity about 45~mK (3~x r.m.s-noise). }\label{fig-all-PI}
\end{figure}

For the interpretation of Galactic polarization emerging from Faraday rotation in the interstellar
medium, also the step of absolute zero-level calibration is required.
Otherwise any interpretation becomes problematic, as it was discussed in some detail by Reich (2006).
Polarization features emerging from Faraday rotation are in general not or very weakly related to total
intensity sources or emission structures. The effect of zero-level calibration is different for
polarized emission, because vectors are added, while for total intensities it is a scalar.

Polarized emission PI and polarization angle $\phi$ are calculated from the observed Stokes parameters U and Q,
assumed to be corrected for all kinds of instrumental effects, by:

\begin{equation}
  \rm PI^{2}~=~U^{2}~+~Q^{2} \quad and \quad  \phi~=~0.5~atan (U/Q)
\end{equation}
Adding missing components in $\rm U_{zero}$ and $\rm Q_{zero}$ gives:

\begin{figure}
\centering
 \includegraphics[bb = 105 105 511 680,width=8cm,angle=270,clip=]{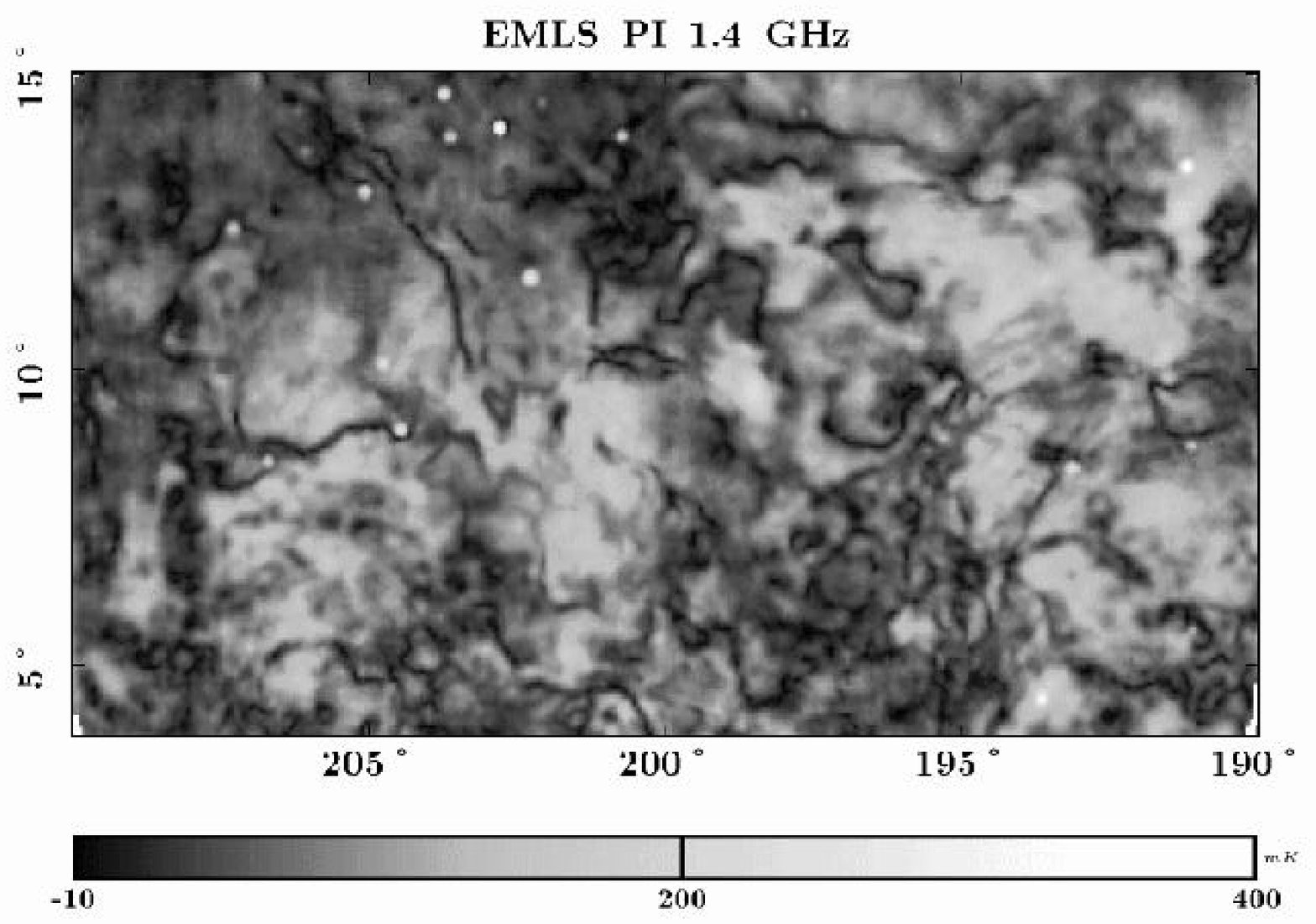}
 \includegraphics[bb = 105 105 511 680,width=8cm,angle=270,clip=]{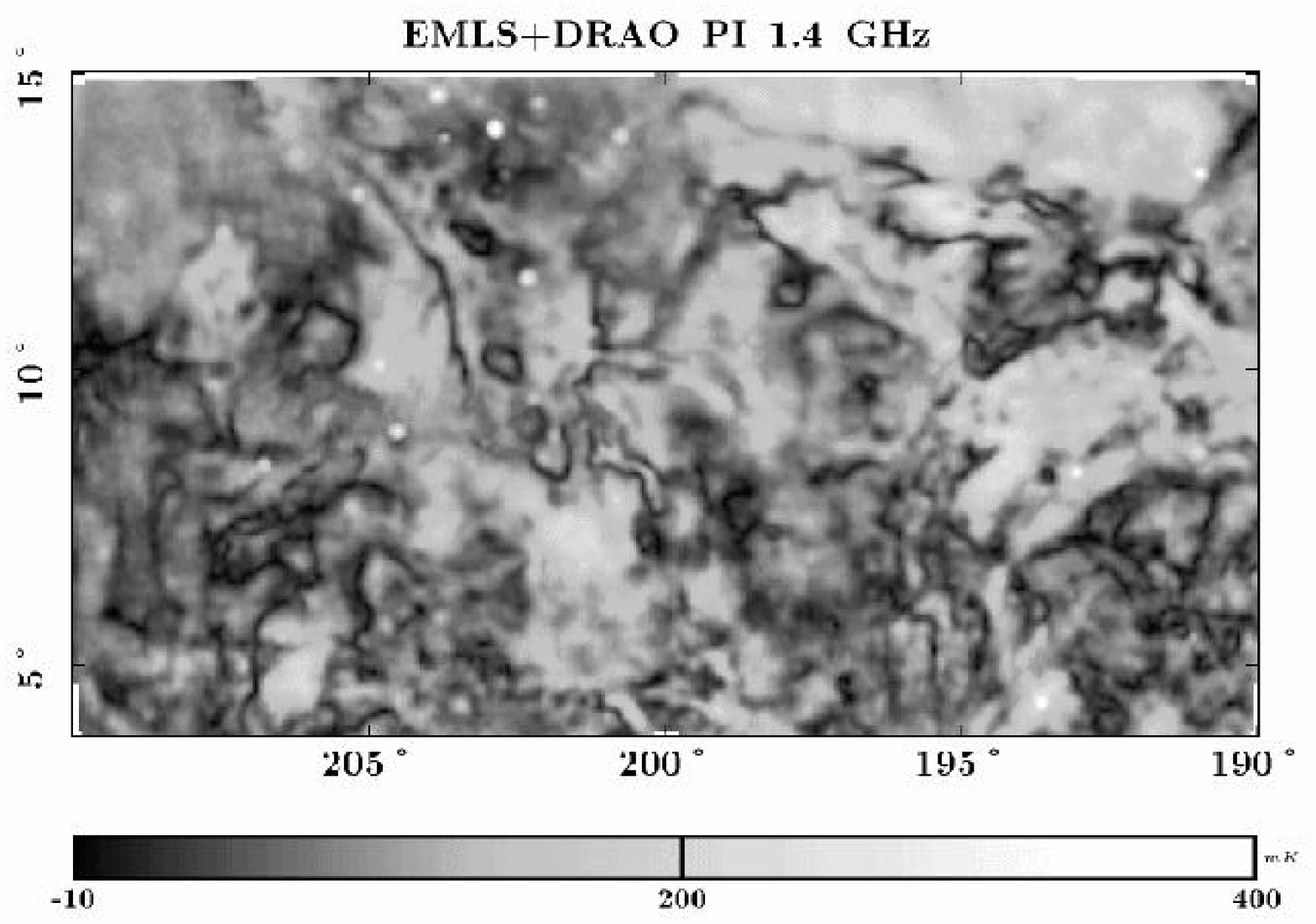}
  \caption{Example map showing polarized intensities from the 1.4~GHz "Effelsberg Medium Latitude Survey"
(Uyan{\i}ker {\it et al.} (1999)) with and without absolute zero-level adjustment. The upper panel shows the
original Effelsberg map of polarized intensities. The lower panel shows the Effelsberg map with added
large-scale components in U and Q exceeding the map size. }\label{fig-anti1}
\end{figure}

\begin{equation}
  \rm PI_{abs}^{2}~=~(U+U_{zero})^{2}~+~(Q+Q_{zero})^{2} \quad   \phi~=~0.5~atan ((U+U_{zero})/(Q+Q_{zero}))
\end{equation}
PI and $\phi$ depend on U(Q) and $\rm U_{zero}(Q_{zero})$  in a non-linear way. U(Q) and $\rm U_{zero}(Q_{zero})$
may have rather different levels and signs, that the inclusion of $\rm U_{zero}(Q_{zero})$ does not only
change the level of PI, but also largely influences the morphology of the observed structures. This is demonstrated
in Fig.~4, where 1.4~GHz polarized emission observed with the Effelsberg 100-m telescope is shown. The numerous
small-scale polarization structures ('canals' and 'rings') have almost no counterpart in the very smooth
total intensity map showing a smooth positive temperature gradient towards the Galactic plane and many
compact extragalactic sources (see Uyan{\i}ker {\it et al.} (1999)). Fig.~4 shows the polarized emission as it was
observed and after a correction for missing large-scale components provided by the northern sky polarization
survey by Wolleben {\it et al.} (2006). The polarization level increases including large-scale components, but
most remarkable is the change of morphology of the small-scale structures, which may change from absorption
into emission structures and vice versa.
Without the large-scale correction any physical interpretation or radiation transfer modelling
of the observed features is strongly limited.

While the WMAP total intensity surveys need a zero-level correction as described above, this does not hold
for the polarization data, where the maps are at a correct zero-level. WMAP observes difference signals
from two feeds with about $140^{\circ}$ viewing angle difference (Bennett {\it et al.} (2003)). Other than for
Stokes I, the Stokes parameter U and Q have no isotropic component.

\subsection{Polarization surveys of the Galactic plane}

With the availability of the 1.4~GHz all-sky polarization survey, Galactic plane surveys at 1.4~GHz
can be properly calibrated as demonstrated in Fig.~4 for a section of the Effelsberg 1.4~GHz survey.
The Effelsberg maps combined with the DRAO 26-m survey were then used to correct the 1.4~GHz polarized
emission component of the CGPS, which results in a Galactic plane survey with 1' angular resolution
showing unprecedented details of the magnetized interstellar medium (Landecker et al., in prep.).
This survey covers the Galactic plane in the longitude range $65^{\circ} \le l \le 175^{\circ}$ and
latitudes between $-3.5^{\circ} \le b \le 5.5^{\circ}$.
The filtering applied to the three surveys before merging the U and Q maps is shown in Fig.~5.

The WMAP polarization survey at 22.8~GHz (Page {\it et al.} (2007)) proves to be very valuable for the correction
of high-resolution polarization data at a few GHz. Faraday rotation of the polarization angles at high latitudes
and towards the anti-centre direction are small, that an extrapolation of the WMAP U and Q maps
from 22.8~GHz towards lower frequencies is possible. This requires the correct spectral index for polarized
emission, which should be close to that observed for total intensity synchrotron emission.
For example a rotation measures
(RM) of 50~rad~$\rm m^{-2}$ causes a polarization angle rotation of about $10^{\circ}$ at $\lambda$6\ cm.
This technique was
applied to the running Sino-German $\lambda$6\ cm (4.8 GHz) polarization survey of the Galactic plane, which is
carried out with the Urumqi 25-m telescope of NAOC/China. As described by Sun {\it et al.} (2007)
this survey aims to cover the Galactic plane for
latitudes between $\pm 5^{\circ}$ with an angular resolution of 9.5' and a rms-sensitivity of
1.4(0.7)~mK~$\rm T_{B}$ for total (polarized) intensities.
Measurements of an absolute polarization level of a few milli-Kelvin is a very ambitious task for a
multi-purpose instrument like the Urumqi 25-m telescope and should be done with dedicated small instruments.
The C-BASS project (Pearson \& C-BASS collaboration 2007)  might be able to provide this required information.

Sun {\it et al.} (2007) published the first survey map from the Urumqi $\lambda$6\ cm survey and demonstrated the
application of the WMAP zero-level correction for an area centered at Galactic longitude $125.5^{\circ}$.
The zero-level offset extrapolated from the smoothed U and Q 22.8~GHz maps was found to be -0.3~mK~$\rm T_{B}$ and
 8.3~mK~$\rm T_{B}$ for U and Q, respectively. The large value for Q indicates that the magnetic field runs
almost parallel to the Galactic plane, as expected.
The observed maps show a lot of diffuse polarized structures partly associated with SNRs, but also in the direction
of some HII-regions, which act as Faraday screens and prove the presence of well-ordered magnetic fields of a
few $\mu$G along the line-of-sight. Of particular interest is a large polarized structure in the field, G125.6-1.8, with a diameter of about 70' showing no signature in total intensity nor in H$\alpha$-emission. This structure changes from a polarized emission structure as seen in the original Urumqi map into an absorption structure after WMAP calibration.
The properties of this feature were modelled by Sun {\it et al.} (2007). For a size of 58~pc and a RM of 200~rad~$\rm m^{-2}$ the calculated magnetic field strength exceeds 6.9~$\mu$G along the line-of-sight for an upper limit
of the thermal electron density of 0.84~cm$^{-3}$ and assumed spherical symmetry. Such regular magnetic
fields  exceed the strength of the regular Galactic magnetic
field, which is about 2$\mu$G in this area according to Han {\it et al.} (2006). The origin of these magnetic bubbles
is not known, and it is also yet unclear how numerous they are and to what extent they influence the properties
of the magnetized interstellar medium. Structures with such a high RM can not be studied at low frequencies,
because small RM fluctuations cause depolarization across the observing beam. G125.6-1.8 is outstanding
at $\lambda$6\ cm, but remains undetected in the $\lambda$21\ cm Galactic plane polarization
survey by Landecker et al. (in prep.).

\begin{figure}
\centering
 \includegraphics[bb = 31 31 470 338,width=10cm,angle=0,clip=]{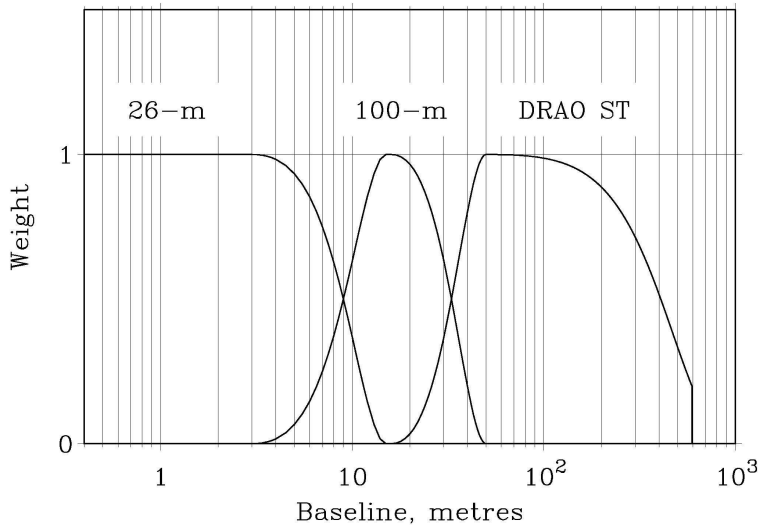}
  \caption{Weighting scheme of spatial structures applied to the three surveys as indicated (Landecker {\it et al.}
in prep.). The combined maps in U and Q result in a Galactic plane survey, which is complete for
all polarized structures larger than 1' (see text).}\label{cov}
\end{figure}

\subsection{Faraday screen observations}

\begin{figure}
\centering
 \includegraphics[bb = 100 200 505 593,width=10cm,angle=270,clip=]{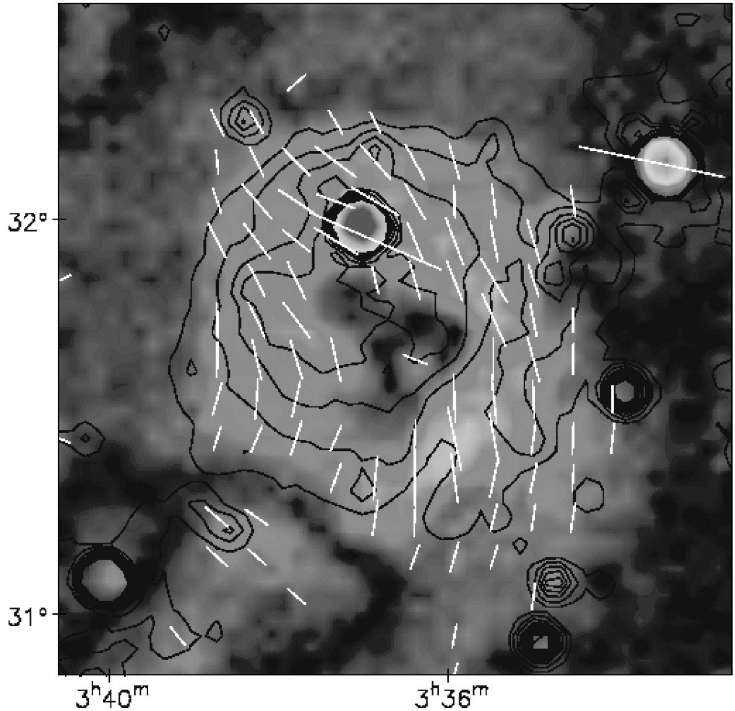}
  \caption{$\lambda$11\ cm Effelsberg observations of the Perseus dust cloud G159.5-18.5 in
total intensity (contours) and polarised intensity (greyscale and vectors).
The linear polarization data indicate strong
emission apparently originating from the thermal gas of G159.5-18.5. This is a clear indication
that G159.5-18.5 acts as a Faraday screen rotating the polarization angle of
background emission (see text).}\label{fig-anti2}
\end{figure}

Numerous Galactic polarized features were already discussed in the literature, but a number
of them clearly suffer from insufficient calibration.
As an example for a misinterpretation of that kind we discuss the case of the dust complex G159.5-18.5
in Perseus (Reich \& Gao, in prep.). G159.5-18.5 is an extended source of about $1^{\circ}$ in size
and very well studied at optical and infrared wavelengths, star light polarization and molecular
line emission (Ridge {\it et al.} (2006) and references therein).

Recently G159.5-18.5 received particular attention, because it is one of the very
few well established cases, where emission from spinning dust particles was clearly detected.
Watson {\it et al.} (2005) presented observations made with the COSMOSOMAS telescope
in the frequency range between 11~GHz and 17~GHz, which were combined with WMAP data
at higher frequencies, showing a spectrum as expected for spinning dust emission, which is
clearly in excess over the weak thermal emission. Battistelli {\it et al.} (2006) presented 11~GHz polarization
observations with the COSMOSOMAS telescope of a large area in Perseus and detected faint polarized emission
of 3.4\% +1.5\%/-1.9\% from G159.5-18.5, which they interpreted as polarized emission emerging from
spinning dust grains, which would be the first detection of that kind. However, the polarization data are
on a relative zero-level in U and Q.

Watson {\it et al.} (2005) extracted information of the thermal emission component of G159.5-18.5 from existing
ground based all-sky surveys. Recent pointed observations with the Effelsberg 100-m telescope
at $\lambda$11~cm and the Urumqi 25-m telescope at $\lambda$6~cm with higher quality clearly confirm the existence
of optically thin thermal gas in G159.5-18.5 with a flux density of about 3.8~Jy, more than four times
below the flux density of spinning dust seen at COSMOSOMAS wavelengths.
Surprisingly strong polarized emission was observed at $\lambda$11~cm (Fig.~6) and
$\lambda$6~cm, which is not expected to originate
from thermal gas. Spinning dust plays no role at $\lambda$11~cm wavelength, that the only valid interpretation
is that G159.5-18.5 acts as a Faraday screen hosting a strong regular magnetic field along the line-of-sight. The
Faraday screen rotates the background polarized emission of G159.5-18.5, which then adds to foreground
polarization in a different way than in the area outside of G159.5-18.5.
In fact, correcting the Battistelli {\it et al.} (2006) polarization data of G159.5-18.5 for missing large-scale
emission as observed by WMAP reduces the polarized emission signal below that of its surroundings.

The Faraday screen model as discussed by Reich \& Gao (in prep.) fits all available polarization observations
with a RM of the order of 190~rad~m$^{-2}$ implying a line-of-sight magnetic field of about 10~$\mu$G.

\section{Conclusions}\label{sec:concl}

We have discussed calibration methods being applied to all-sky and Galactic plane continuum surveys
in total intensity and linear polarization. The interpretation of polarized emission emerging from
Faraday rotation effects in the interstellar medium requires special attention, in paparticularn
absolute zero-level setting is essential.

\begin{acknowledgments}

We like to thank all our colleagues at MPIfR/Bonn, IAR/Argentina, DRAO/Canada and NAOC/China for
their invaluable contributions to the various surveys projects discussed in this paper.

\end{acknowledgments}

\end{document}